\documentstyle[array,12pt]{article}
\def\MC{{\cal M}}

\begin{document}

\baselineskip=0.6cm

\noindent P.~N.~Lebedev Institute Preprint     \hfill
FIAN/TD/18--92\\ I.~E.~Tamm Theory Department       \hfill
\begin{flushright}{November 1992}\end{flushright}

\begin{center}

\vspace{0.5in}

{\Large\bf On possible generalizations of field--antifield formalism}

\bigskip

\vspace{0.3in}
{\large  I.~A.~Batalin and I.~V.~Tyutin}\\
\medskip  {\it Department of Theoretical Physics} \\ {\it  P.~N.~Lebedev
Physical Institute} \\ {\it Leninsky prospect, 53, 117 924, Moscow,
Russia}$^{\dagger}$\\

\end{center}

\vspace{1.5cm}

\centerline{\bf ABSTRACT}
\begin{quotation}

A generalized version is proposed for the field--antifield formalism.
The antibracket operation is defined in arbitrary field--antifield
coordinates. The antisymplectic definitions are given for first-- and
second--class constraints. In the case of second--class constraints the
Dirac's antibracket operation is defined. The quantum master equation
as well as the hypergauge fixing procedure are formulated in a
coordinate--invariant way. The general hypergauge functions are shown
to be antisymplectic first--class constraints whose Jacobian matrix
determinant is constant on the constraint surface. The BRST--type
generalized transformations are defined and the functional integral
is shown to be independent of the hypergauge variations admitted. In
the case of reduced phase space the Dirac's antibrackets are used
instead of the ordinary ones.

\end{quotation}
\vfill
\noindent
$^{\dagger}$ E-mail address: lirc@glas.apc.org
\newpage
\pagestyle{plain}

\section{Introduction}

Covariant quantization of gauge--field systems has a long--time history
started from the famous works of Feynman \cite{1}, Faddeev and Popov \cite{2}
and DeWitt \cite{3}.

A unique closed approach to the covariant quantization problem has
been proposed in work \cite{4} of Batalin and Vilkovisky. These authors
have introduced the field--antifield phase space concept as well as
the antibracket operation that is an antisymplectic counterpart of
the well--known Poisson bracket. Moreover, a nilpotent second--order
differential operator has been discovered, that differentiates the
antibracket according to the Leubnitz rule. Due to the mentioned
property, henceforth we shall refer this remarkable operator as
``antisymplectic differential'', although this terminology does not
coincide with the standard one of the exterior form theory.

The authors of the paper \cite{4} have formulated the general quantization
principle to be applied directly to the Lagrangian formalism. The
principle requires for the exponential of $i/\hbar$ times quantum action
to be annihilated by the antisymplectic differential. Thus the quantum
master equation has appeared to acquire its great importance. The
corresponding classical master equation requires for the classical
master action to commute with itself in the antibracket sense to give
zero.

In its own turn, the classical master equation defines an universal gauge
hypertheory whose hypergauge generators are always nilpotent
at the classical hyperextremals. The classical master action, that
possesses a minimally--possible hypergauge degeneracy, is called
``proper''. This minimal degeneracy is removed exactly by imposing the
standard BV hypergauge conditions that require for all the antifields
to equal to the corresponding field derivatives of a Fermionic
function. Due to the quantum master equation, the functional integral
does not depend formally on the hypergauge Fermionic function
variations. If one uses the proper master action together with the
standard BV hypergauge, then the functional integral is certainly
nondegenerate and thus calculable via the loop expansion technique.

A general strategy of the BV approach is to involve a given gauge
theory into the universal hypertheory that is determined by the
proper master action. Moreover, the necessary spectrum of the field--antifield
variables is determined in such a way that just provides
for the master action to be proper. That is the mechanism by means
of which all the ghost generations appear naturally.

The above--mentioned strategy has been applied successfully to the gauge
theories with irreducible open algebras \cite{4} and to the theories with
linearly--dependent gauge generators \cite{5}, as well. Also the recent
developments \cite{6,7,8,22} in secondary--quantized string field theory are
substantially based on the BV approach.

Many authors have contributed to develop and apply the field--antifield
formalism. For detailed references see the review lecture of Henneaux \cite{9}.

The contributions of Zinn--Justin \cite{10}, Kallosh \cite{11}, de Wit and
van Holten \cite{12} had been important to reveal the general status of the
classical master equation.

Witten \cite{13} has given a deep geometric interpretation of the quantum
master equation.

An Sp(2)--covariant version of the BV formalism has been proposed
recently by Batalin, Lavrov and Tyutin \cite{14,15,16}.

Henneaux \cite{17} has extended the Witten's interpretation to cover the
Sp(2)--covariant formulation.

A relation between the Hamiltonian BFV and Lagrangian BV formalisms
has been revealed by Grigoryan, Grigoryan and Tyutin \cite{18}. These
authors have used a functional counterpart of the operator method
proposed originally by Batalin and Fradkin \cite{19}.

Independently of the gauge field quantization problem, an invariant geometric
description of the symplectic and antisymplectic structures on the K\"ahlerian
superspaces has been given by Khudaverdian and Nersessian \cite{23,24}.

Volkov et all \cite{25,26} studied the antibracket reformulation and
quantization of supersymmetric mechanics.

In the present work we undertake further steps in developing the
field--antifield formalism.

The first problem is to give a coordinate--invariant formulation to the
quantum master equation and to the hypergauge fixing procedure as well.

The second problem is to assign the antifields to the hypergauge
Lagrangian multipliers and thus to give start to the hierarchical
proliferation process that introduces the hypergauges of higher levels.

The third problem is to define the Dirac's counterpart of the field--antifield
formalism in case of the basic field--antifield variables
reduced originally by second--class constraints.

The paper is organized as follows.

In Section 2 we define in a coordinate--invariant way the Fermionic
generating operator to be nilpotent. The nilpotency condition gives
automatically the integrability property of the connection field,
together with the equation for the measure density and the
antisymplecticity property of the phase space metric.

Having the antisymplectic metric, we define the antibracket operation
in a coordinate--invariant way. The above--mentioned nilpotent generating
operator differentiates the invariant antibracket by the Leubnitz rule.
It is quite evident that this operator is nothing other but the
antisymplectic differential in its coordinate--invariant version.

In terms of the antibracket operation we give the antisymplectic
definitions for constraints to be of the first or second class. Thus
we formulate the antibracket involution relations to be a definition
of first--class constraints. In case of second--class constraints we
define the antisymplectic counterpart of the well--known Dirac's
bracket. Then we formulate the equation for the Dirac's measure
density of the reduced field--antifield phase space.

In Section 3 we define and consider in details the general invariant
form for the Lagrangian functional integral of the first level. Being
of the first level, this functional integral, by definition, contains
the hypergauge conditions imposed directly on the basic field--antifield
variables.

We formulate the quantum master equation as well as the hypergauge
fixing procedure in a coordinate--invariant way.

Instead of the standard BV hypergauge conditions, we formulate the
unimodular involution relations that require for the hypergauge
functions to be antisymplectic first--class constraints under the
extra conditions that control the constraint Jacobian matrix
determinant.

Then we define the BRST--type generalized transformations to show the
functional integral to be independent formally of the hypergauge
function variations admitted.

We consider in details the conditions that provide for the hypergauge
functions to remove a degeneracy of the functional integral. Thus we
confirm that the functional integral is certainly nondegenerate if one
uses the proper master action together with the hypergauge admitted.

Also, we show that the natural arbitrariness of the phase space
measure density can be absorbed into redefinition of the quantum
action, whereas its classical part remains unchanged.

In Section 4 we give start to the hierarchical proliferation process
that introduces the higher level hypergauges. Actually, we consider
here the functional integral of the second level only.

To begin with, we assign their own antifields to the hypergauge
Lagrangian multipliers of the first level. Thus we extend the original
field--antifield phase space by including the new anticanonical pairs.

In the extended phase space we formulate the quantum master equation
for the second--level quantum action. Then we impose the second--level
hypergauge conditions on the new antifields assigned to the first--level
Lagrangian multipliers. The second--level Lagrangian multipliers
do not possess at this stage their own antifields to appear at the
third level and so on.

The quantum master equation of the second level appears to be a
generating mechanism for the first--level unimodular involution
relations that follow to the lowest order in new field--antifield
pairs. This generating mechanism in a remarkable way synthesizes
in itself the characteristic features of the Hamiltonian and
Lagrangian gauge algebra generating equations, whereas, the new
anticanonical pairs play the role of the Hamiltonian ghost variables.

In their own turn, the second--level hypergauge functions are
subordinated to the new unimodular involution equations to be called
``second--level'' ones.

After the second--level functional integral is constructed, we realize
completely the corresponding counterpart of the program undertaken
in the previous Section. Particularly, we show that the second--level
functional integral depends actually neither on the first-- nor on
the second--level hypergauge function variations admitted.

In Section 5 we comment in brief the modifications needed for the
above--considered formalism in case of the field--antifield phase
variables reduced originally by antisymplectic second--class
constraints. as an example we give an explicit form for the Dirac's
counterpart of the first--level functional integral.

Notation and Convention. as is usual, $\varepsilon(A)$ denotes the
Grassmann parity of a quantity $A$. By $\hbox{rank}||X_{AB}||$ we denote
maximal size of the invertible square block of a supermatrix $||X_{AB}||$.

Other notation is clear from the context.

\section{Antisymplectic Differential and Antibrackets}

Let:

$$\Gamma^A,\quad A=1,\ldots,2N,\quad\varepsilon(\Gamma^A)\equiv\varepsilon_A,
\eqno{(2.1)}$$
be a set of field--antifield variables:

$$\{\Gamma^A\}=\{\varphi^a,\varphi^*_a|a=1,\ldots,N,\varepsilon(\varphi^*_a)=
\varepsilon(\varphi^a)+1\}.\eqno{(2.2)}$$
We consider the variables (2.2) to be local coordinates of the
corresponding field--antifield phase space $\MC$.

Let $E^{AB}(\Gamma)$ be a nondegenerate odd contravariant metric with the
adjoint--antisymmetry property:

$$\varepsilon(E^{AB})=\varepsilon_A+\varepsilon_B+1,\quad E^{AB}=
-E^{BA}(-1)^{(\varepsilon_A+1)(\varepsilon_B+1)},\eqno{(2.3)}$$
while $F_A(\Gamma)$, $\varepsilon(F_A)=\varepsilon_A$, be a ``connection''
field.

Let us introduce the following generating operator
\footnote{In fact, the operator (2.4) is nothing other but the general form of
a
second--order differential operator whithout the derivativeless term. From this
viewpoint even the adjoint--antisymmetry property (2.3) is not to be imposed
imperatively. When being nonzero, the adjoint--symmetric part of $E^{AB}$ can
be absorbed into redefinition of the ``connection'' field $F$.}:

$$\Delta\equiv{1\over2}(-1)^{\varepsilon_A}(\partial_A+F_A)E^{AB}\partial_B,
\quad\varepsilon(\Delta)=1,\eqno{(2.4)}$$
to be nilpotent:

$$\Delta^2=0\eqno{(2.5)}$$

This nilpotency condition gives immediately:

$$\Delta(-1)^{\varepsilon_C}(\partial_C+F_C)E^{CD}=0,\eqno{(2.6)}$$

$$\partial_AF_B-\partial_BF_A(-1)^{\varepsilon_A\varepsilon_B}=0,
\eqno{(2.7)}$$

$$(-1)^{(\varepsilon_A+1)(\varepsilon_C+1)}E^{AD}\partial_DE^{BC}+
\hbox{cycle}(A,B,C)=0.\eqno{(2.8)}$$

The equation (2.7) gives locally:

$$F_A=\partial_A\ln M(\Gamma),\eqno{(2.9)}$$
so that the equation (2.6) determines, in fact, the scalar density
$M(\Gamma)$:

$$\Delta(-1)^{\varepsilon_C}M^{-1}\partial_CME^{CD}=0,\eqno{(2.10)}$$

$$\Delta={1\over2}(-1)^{\varepsilon_A}M^{-1}\partial_AME^{AB}\partial_B.
\eqno{(2.11)}$$
An invariant measure $d\mu(\Gamma)$ on the phase space $\MC$:

$$d\mu(\Gamma)\equiv M(\Gamma)d\Gamma\eqno{(2.12)}$$
is naturally associated with the density $M(\Gamma)$.

In its own turn, the cyclic equation (2.8) is nothing other but the
antisymplecticity property of the metric $E^{AB}$. This property allows
one to introduce naturally the antibracket operation:

$$(A,B)\equiv A\overleftarrow{\partial_C}E^{CD}\overrightarrow{\partial_D}B
\eqno{(2.13)}$$
with the following algebraic properties \cite{21,4}:

$$\varepsilon\left((A,B)\right)=\varepsilon(A)+\varepsilon(B)+1,\eqno{(2.14)}$$

$$(A,B)=-(B,A)(-1)^{(\varepsilon(A)+1)(\varepsilon(B)+1)},
\eqno{(2.15)}$$

$$(A,BC)=(A,B)C+B(A,C)(-1)^{(\varepsilon(A)+1)\varepsilon(B)},
\eqno{(2.16)}$$

$$\left((A,B),C\right)(-1)^{(\varepsilon(A)+1)(\varepsilon(C)+1)}+
\hbox{cycle}(A,B,C)=0.\eqno{(2.17)}$$
Besides, the formula \cite{20}:

$$\Delta(A,B)=(\Delta A,B)+(A,\Delta B)(-1)^{\varepsilon(A)+1}.
\eqno{(2.18)}$$
represents the property of the antibracket operation with respect to
applying the operator $\Delta$.

On the other hand, applying the operator $\Delta$ to the ordinary product
$AB$, we have:

$$\Delta(AB)=(\Delta A)B+(A,B)(-1)^{\varepsilon(A)}+
A(\Delta B)(-1)^{\varepsilon(A)}.\eqno{(2.19)}$$

The formula (2.18) shows that the operator $\Delta$ differentiates the
antibracket $(A,B)$ according to the Leubnitz rule. Due to this remarkable
property, the nilpotent operator $\Delta$ is called ``antisymplectic
differential''.

Having the antibracket operation at our disposal, we can introduce
naturally a formal counterpart of the well--known Dirac's terminology
in order to define constraints to be of the first or second class.

By definition, the functions:

$$G_i(\Gamma),\quad i=1,\ldots,K,\quad\varepsilon(G_i)\equiv\varepsilon_i.
\eqno{(2.20)}$$
are called first--class constraints if the antibracket involution relations
hold:

$$(G_i,G_j)=G_kU^k_{ij},\eqno{(2.21)}$$
where the structural coefficients $U^k_{ij}(\Gamma)$ possess the properties:

$$\varepsilon(U^k_{ij})=\varepsilon_i+\varepsilon_j+\varepsilon_k+1,
\quad U^k_{ij}=-U^k_{ji}(-1)^{(\varepsilon_i+1)(\varepsilon_j+1)}.
\eqno{(2.22)}$$

Alternatively, the functions:

$$\Theta_\alpha(\Gamma),\quad\alpha=1,\ldots,2L,\quad\varepsilon(\Theta_\alpha)
\equiv\varepsilon_\alpha.\eqno{(2.23)}$$
are called second--class constraints if their antibracket matrix:

$$Q_{\alpha\beta}(\Gamma)\equiv(\Theta_\alpha(\Gamma),\Theta_\beta(\Gamma))
\eqno{(2.24)}$$
is nondegenerate:

$$\exists\,Q^{\alpha\beta}\,:\quad Q_{\alpha\beta}Q^{\beta\gamma}=
{\delta_\alpha}^\gamma,\eqno{(2.25)}$$
so that we have:

$$\varepsilon(Q^{\alpha\beta})=\varepsilon_\alpha+\varepsilon_\beta+1,\quad
Q^{\alpha\beta}=-Q^{\beta\alpha}(-1)^{\varepsilon_\alpha\varepsilon_\beta}.
\eqno{(2.26)}$$

If the condition (2.25) is satisfied, then one can define the following
antisymplectic counterpart of the well--known Dirac's bracket:

$$(A,B)_{(D)}\equiv(A,B)-(A,\Theta_\alpha)Q^{\alpha\beta}(\Theta_\beta,B).
\eqno{(2.27)}$$

It can be checked directly that the Dirac's anti--bracket (2.27) possesses
all the algebraic properties (2.14)-(2.17).

Having the anti--bracket (2.27), one can define naturally the Dirac's
antisymplectic differential:

$$\Delta_{(D)}\equiv{1\over2}(-1)^{\varepsilon_A}M^{-1}_{(D)}
\partial_AM_{(D)}E^{AB}_{(D)}\partial_B,\quad E^{AB}_{(D)}\equiv
(\Gamma^A,\Gamma^B)_{(D)},\eqno{(2.28)}$$
to be nilpotent:

$$\Delta^2_{(D)}=0,\eqno{(2.29)}$$
that gives the following equation for the density $M_{(D)}(\Gamma)$:

$$\Delta_{(D)}(-1)^{\varepsilon_A}M^{-1}_{(D)}
\partial_AM_{(D)}E^{AB}_{(D)}=0.\eqno{(2.30)}$$

The corresponding Dirac's measure has the form:

$$d\mu_{(D)}(\Gamma)\equiv M_{(D)}(\Gamma)\delta(\Theta)d\Gamma.
\eqno{(2.31)}$$

Of course, the Dirac--type counterparts of the equations (2.18), (2.19)
hold true for the antibracket (2.27) and differential (2.28).

Concluding this Section, we have to make the following remark.

Contrary to the standard symplectic formalism, in the present,
antisymplectic, case one cannot express the measure densities $M(\Gamma)$ or
$M_{(D)}(\Gamma)$ explicitly (i.e. algebraically) in terms of the metric
$E^{AB}$ or, respectively, of the metric $E^{AB}_{(D)}$ and the matrix
$Q_{\alpha\beta}$. To a considerable extent, this is because of the fact
that the standard superdeterminant concept does not work for odd supermatrices
such as the ones $E^{AB}$ or $Q_{\alpha\beta}$.

\section{ General Functional Integral of the First Level}

The antisymplectic differential concept plays the key role when
constructing the general form for the Lagrangian functional integral.
In fact, the basic idea of the BV--approach is to involve an initially--given
gauge theory into the universal hypertheory whose hypergauge
generators are always nilpotent at the classical hyperextremals. To
define the above--mentioned hypertheory in a most natural and effective
way, one should require for the exponential of $i/\hbar$ times quantum action
to be annihilated by the antisymplectic differential $\Delta$. Thus we
arrive at the quantum master equation.

To define the functional integral to be nondegenerate, one needs
hypergauge fixing. If the hypergauge conditions are imposed directly
on the basic field--antifield variables (2.1), then, by definition,
the functional integral is called ``of the first lel'', and so on.

So, we propose the following basic formula for the general Lagrangian
functional integral of the first level:

$$Z=\int\!\!\exp\{{i\over\hbar}[W(\Gamma)+G_a(\Gamma)\pi^a]\}d\pi d\mu(\Gamma),
\eqno{(3.1)}$$
where the quantum action $W(\Gamma)$ satisfies the master equation:

$$\Delta\exp\{{i\over\hbar}W\}=0, \eqno{(3.2)}$$
or equivalently:

$${1\over2}(W,W)=i\hbar\Delta W, \eqno{(3.3)}$$
while the hypergauge functions:

$$G_a(\Gamma),\quad a=1,\ldots,N,\quad\varepsilon(G_a)=\varepsilon_a,
\eqno{(3.4)}$$
are subjected to the conditions:

$$(G_a,G_b)=G_cU^c_{ab}, \eqno{(3.5)}$$

$$\Delta G_a-U^b_{ba}(-1)^{\varepsilon_b}=G_bV^b_a, \eqno{(3.6)}$$

$$V^a_a=G_a\tilde{G}{}^a, \eqno{(3.7)}$$
and should remove a gauge degeneracy of the action $W(\Gamma)$.

as for the structure functions:

$$U^c_{ab}=-U^c_{ba}(-1)^{(\varepsilon_a+1)(\varepsilon_b+1)},
\quad V^b_a,\quad \tilde{G}{}^a, \eqno{(3.8)}$$
they are subjected to the compatibility conditions of the equations
(3.5)-(3.7),
only.

The equations (3.5) have the form of the antibracket involution
relations (2.21). Thereby the admitted hypergauge functions $G_a$
appear to be, in fact, first--class constraints. Moreover, these
constraints are restricted by the additional equations (3.6), (3.7)
that control the Jacobian matrix determinant of the functions $G_a$.
Henceforth we shall refer the equations (3.5)-(3.7) as ``unimodular
involution relations''.

It is relevant to note here that the Abelian case $(U^c_{ab}=0)$ of the
relations (3.5) has been proposed in Ref.\cite{5}, but without the
corresponding
unimodularity condition $\Delta G_a=0$.

It is a crucial circumstance that the total set of equations (3.2),
(3.5)-(3.7) provides for the functional integral (3.1) to be invariant
under the BRST-type transformations:

$$\delta\Gamma^A=(\Gamma^A,W-G_a\pi^a)\mu, \eqno{(3.9)}$$

$$\delta\pi^a=(U^a_{bc}\pi^c\pi^b(-1)^{\varepsilon_b}-2i\hbar V^a_b\pi^b-
2(i\hbar)^2\tilde{G}{}^a)\mu,\eqno{(3.10)}$$
where $\mu$ is a Fermionic parameter.

Choosing the parameter $\mu$ to be the function:

$$\mu={i\over2\hbar}\delta\Psi(\Gamma), \eqno{(3.11)}$$
that satisfies the equations:

$$i\hbar\Delta\delta\Psi=G_a\delta K^a,\quad\Delta(G_a\delta K^a)=0,
\eqno{(3.12)}$$
and making the additional variations:

$$\delta\Gamma^A=(\Gamma^A,\delta\Psi),\quad\delta\pi^a=\delta K^a,
\eqno{(3.13)}$$
one generates the following effective change of the hypergauge
functions $G_a$ alone:

$$G_a(\Gamma)\quad\rightarrow\quad G_a\left(\Gamma+
(\Gamma,\delta\Psi)\right), \eqno{(3.14)}$$
in the functional integral (3.1).

Thus, it is proven formally that the functional integral (3.1) does
not depend on the hypergauge variations of the canonical form:

$$\delta G_a=(G_a,\delta\Psi). \eqno{(3.15)}$$

The variations (3.15) certainly retain the form of the unimodular
involution  relations (3.5)-(3.7), but the most general hypergauge
variations with the mentioned property are of the form:
$\delta G_a=(G_a,\delta\Psi)+G_b\delta\Lambda^b_a$. Hence the variation
(3.15) induce the most general actual changes admitted for the hypergauge
surface $G_a=0$. Thus the canonical hypergauge variations (3.15) are shown
to be quite sufficient for our purposes (see also Eq.(3.54)).

Now, following the method of Ref.\cite{4}, let us consider the conditions
that provide for the hypergauge functions $G_a$ to remove a degeneracy
of the classical action. Let us seek for a solution to the equation
(3.3) in the form of $\hbar$--power series expansion:

$$W=S+i\hbar W_1+\ldots\,\,. \eqno{(3.16)}$$

We have:

$$(S,S)=0, \eqno{(3.17)}$$

$$(S,W_1)=\Delta S, \eqno{(3.18)}$$
and so on.

The classical master equation (3.17) just determines the universal
hypertheory with nilpotent hypergauge generators. To see this, let us
differentiate the equation (3.17) to find the N\"other identities:

$${R_A}^B\partial_BS=0, \eqno{(3.19)}$$
where the hypergauge generators are:

$${R_A}^B\equiv[2(\overrightarrow{\partial_A}S\overleftarrow{\partial_C})
+S\overrightarrow{\partial_A}\overleftarrow{\partial_C}]E^{CB}.\eqno{(3.20)}$$
Differentiating the identities (3.19) in their own turn, we find the desired
nilpotency property:

$${R_A}^B{R_B}^C\bigl|_{\partial S=0}=0, \eqno{(3.21)}$$
that gives:

$$\hbox{rank}||{R_A}^B||\bigl|_{\partial S=0}\le N,
\eqno{(3.22)}$$

Next, let us write down the total classical action that enters
the functional integral (3.1):

$$\hbox{Classical action}=S+G_a\pi^a. \eqno{(3.23)}$$
The corresponding motion equations are:

$$\partial_as+(\partial_AG_a)\pi^a=0,\quad G_a=0, \eqno{(3.24)}$$
that give:

$${R_A}^B(\partial_BG_a)\pi^a=0, \eqno{(3.25)}$$
due to the identities (3.19).

A solution to the equations (3.25) for the Lagrangian multipliers $\pi^a$
is unique iff the following uniqueness condition is satisfied:

$$\hbox{rank}||{R_A}^B\partial_BG_a||\bigl|_{\partial S=G=0}=N,
\eqno{(3.26)}$$
so that we have:

$$\hbox{rank}||{R_A}^B||\bigl|_{\partial S=G=0}\ge N,
\eqno{(3.27)}$$

$$\hbox{rank}||\partial_BG_a||\bigl|_{\partial S=G=0}=N.
\eqno{(3.28)}$$

The conditions (3.22) and (3.27) are compatible with each other iff the
equality:

$$\hbox{rank}||{R_A}^B||\bigl|_{\partial S=G=0}=N,
\eqno{(3.29)}$$
holds.

Let the condition (3.28) be satisfied, so that the square $N\times N$--matrix:

$$   ||\partial_{\tilde{B}}G_a||\bigl|_{\partial S=G=0},
\quad\{\tilde{B}\}\subset\{B\},\eqno{(3.30)}$$
is nondegenerate. In order to provide the standard transformation property
for the hypergauge $\delta$--function $\delta(G)$, one should require
for the matrix (3.30) to be even:

$$\tilde{\varepsilon}_b+\varepsilon_a=\bar{\varepsilon}_b+\bar{\varepsilon_a},
\eqno{(3.31)}$$
where $\tilde{\varepsilon}_b$, $b=1,\ldots,N$, denote the parities
$\varepsilon_{\tilde{B}}$ to be naturally ordered, while
$\bar{\varepsilon}_b$ are new parities to be determined by the equation
(3.31). We have two possible solutions:

$$\bar{\varepsilon}_a=\varepsilon_a,\quad\tilde{\varepsilon}_b=
\varepsilon_b, \eqno{(3.32)}$$

$$\bar{\varepsilon}_a=\varepsilon_a+1,\quad\tilde{\varepsilon}_b=
\varepsilon_b, \eqno{(3.33)}$$

as a next step, let us show both solutions (3.32) to be acceptable.
To see this, let us note that the basic equations (2.10), (3.2) admit
a natural arbitrariness for their solutions. First, let us consider
the equation (2.10) for the measure density $M(\Gamma)$. Let $M(\Gamma)$ be a
solution to this equation. Then the function $M(\Gamma)J(\Gamma)$ satisfies the
same equation if the function $J(\Gamma)$ possesses the property:

$$\Delta\sqrt{J}=0. \eqno{(3.34)}$$
Next, let us change the measure density in the functional integral (3.1), as
well as in the equations (3.2), (3.6), according to the rule:

$$M(\Gamma)\quad\rightarrow\quad M(\Gamma)J(\Gamma), \eqno{(3.35)}$$
where the function $J(\Gamma)$ satisfies the equation (3.34). It can be
checked directly that the change (3.35) induces the following transformation
for
the solution W to the equation (3.2):

$$W(\Gamma)\quad\rightarrow\quad W(\Gamma)-
{\hbar\over i}\ln\sqrt{J(\Gamma)}.\eqno{(3.36)}$$
To compensate the changes (3.35), (3.36) in the functional integral (3.1), the
hypergauge $\delta$--function $\delta(G)$ should behave as:

$$\delta(G)\quad\rightarrow\quad\delta(G)(\sqrt{J})^{-1}. \eqno{(3.37)}$$
By making use of the unimodular involution relations (3.5)-(3.7), one can
confirm that the hypergauge $\delta$--function changes under the
transformation (3.35) just according to the rule (3.37) if one chooses either
the solution (3.32) or the one (3.33), so that in fact both these solutions
appear to be effectively equivalent.

So, we have studied all the required conditions  for the hypergauge
functions $G_a$ and thus we have found the unimodular involution relations
(3.5)-(3.7) together with the admissibility conditions (3.28), (3.31).
We have established also that the natural arbitrariness of the measure
density $M(\Gamma)$ can be absorbed certainly into the change (3.36) of the
quantum action $W(\Gamma)$, whereas its classical part S remains unchanged.

Let us return now to the condition (3.29). According to the terminology
of Ref.\cite{4}, the solution S, that satisfies the classical master equation
(3.17) and the condition (3.29), is called ``proper''. If one uses the
proper solution S, whereas the hypergauge functions $G_a$ satisfy the
above--mentioned conditions, then the functional integral (3.1) is
certainly nondegenerate and thus calculable effectively via the loop
expansion technique.

A general strategy of the BV quantization method is to determine a
spectrum of the field--antifield variables $\Gamma^A$ in such a way that just
provides for the master action S to be a proper solution. That is the
mechanism by means of which all the ghost generations appear naturally.
This strategy has been applied successfully to the irreducible
theories with open gauge algebras and to the theories with linearly-
dependent gauge generators, as well.

To conclude this Section, let us demonstrate explicitly that the standard
version of the BV--formalism \cite{4} follows directly from the general
functional representation (3.1) if one chooses the Darboux coordinates:

$$E^{AB}=\left(\begin{array}{cc}0&\delta^{ab}\\-\delta^{ab}&0\end{array}\right),
\quad(A,B)=A(\overleftarrow{\partial_a}\overrightarrow{\partial_*^a}-
\overleftarrow{\partial_*^a}\overrightarrow{\partial_a})B,\eqno{(3.40)}$$
and the trivial measure density:

$$M=1,\quad\Delta=(-1)^{\varepsilon(\varphi^a)}\partial_a\partial_*^a.
\eqno{(3.41)}$$

Let the hypergauge functions $G_a$ be explicitly solvable with respect
to the antifield variables $\varphi^*_a$:

$$G_a(\varphi,\varphi^*)=(\varphi^*_b-f_b(\varphi))
\Lambda^b_a(\varphi,\varphi^*), \eqno{(3.42)}$$
where $\Lambda^b_a(\varphi,\varphi^*)$ is an even nondegenerate matrix.

Choosing the solution (3.33), we have:

$$\varepsilon(\Lambda^b_a)=\bar{\varepsilon}_a+\bar{\varepsilon}_b,
\eqno{(3.43)}$$
where:

$$\bar{\varepsilon}_a=\varepsilon_a+1=\varepsilon(\varphi^*_a)+1=
\varepsilon(\varphi^a). \eqno{(3.44)}$$

Substituting the ansatz (3.42) into the antibracket involution relations
(3.5), we have at $\varphi^*=f(\varphi)$:

$$(\varphi^*_a-f_a(\varphi),\varphi^*_b-f_b(\varphi))=0, \eqno{(3.45)}$$
that gives locally:

$$f_a(\varphi)=\partial_a\Psi(\varphi),\quad\varepsilon(\Psi)=1.
\eqno{(3.46)}$$

Let us substitute this solution into the ansatz (3.42) and then expand
the result near the hypersurface $\varphi^*_a=\partial_a\Psi(\varphi)$:

$$
\begin{array}{c}G_a(\varphi,\varphi^*)=(\varphi^*_b-
\partial_b\Psi(\varphi))\Lambda^b_a(\varphi,\varphi^*)=\\[9pt]
=(\varphi^*_b-\partial_b\Psi(\varphi))\Lambda^b_a(\varphi)+{1\over2}
(\varphi^*_c-\partial_c\Psi(\varphi))(\varphi^*_b-\partial_b\Psi(\varphi))
\Lambda^{bc}_a(\varphi)+\ldots,\end{array} \eqno{(3.47)}$$
where:

$$\Lambda^b_a(\varphi)\equiv\Lambda^b_a(\varphi,\varphi^*
=\partial\Psi(\varphi)),\eqno{(3.48)}$$

$$\Lambda^{bc}_a(\varphi)=\Lambda^{cb}_a(\varphi)
(-1)^{(\bar{\varepsilon}_b+1)(\bar{\varepsilon}_c+1)}. \eqno{(3.49)}$$

To the first order in $\left(\varphi^*-\partial\Psi(\varphi)\right)$ the
involution relations (3.5) give:

$$\Lambda^c_a(\varphi)\overleftarrow{\partial_d}\Lambda^d_b(\varphi)-
\Lambda^c_b(\varphi)\overleftarrow{\partial_d}\Lambda^d_a(\varphi)
(-1)^{\bar{\varepsilon}_a\bar{\varepsilon}_b}=\Lambda^c_d(\varphi)
U^d_{ab}(\varphi),\eqno{(3.50)}$$
where:

$$U^c_{ab}(\varphi)\equiv U^c_{ab}(\varphi,\varphi^*=\partial\Psi(\varphi)).
\eqno{(3.51)}$$

In its own turn, the equation (3.6) gives at
$\varphi^*=\partial\Psi(\varphi)$:

$$(-1)^{\bar{\varepsilon}_b}\partial_b\Lambda^b_a(\varphi)=-
U^b_{ba}(\varphi)(-1)^{\bar{\varepsilon}_b}. \eqno{(3.52)}$$
It should be noted here that even the second term in (3.47) does not contribute
to $\Delta G_a$ at $\varphi^*=\partial\Psi(\varphi)$ because of the
adjoint--symmetry property (3.49).

It follows from the equations (3.50), (3.52) that:

$$\partial_a\ln\det\Lambda(\varphi)=0, \eqno{(3.53)}$$
and hence:

$$\det\left(\partial_*G(\varphi,\varphi^*)\right)\bigl|_{\varphi^*=
\partial\Psi(\varphi)}=\hbox{const}.\eqno{(3.54)}$$
This constancy property is an explicit example how the equation (3.6) controls
the Jacobian matrix determinant of the hypergauge functions $G_a$, and that
is why the equations (3.5)-(3.7) are called ``the unimodular involution
relations''.

Due to the property (3.54), we find:

$$\delta(G)=\hbox{const}\cdot\delta\left(\varphi^*-\partial\Psi(\varphi)\right),
\eqno{(3.55)}$$
that is the standard BV--gauge.

Finally, we obtain the standard functional integral \cite{4}:

$$Z_{standard}=\int\!\!\exp\{{i\over\hbar}W\left(\varphi,\varphi^*=
\partial\Psi(\varphi)\right)\}d\varphi, \eqno{(3.56)}$$

$$\overrightarrow{\partial^a_*}\exp\{{i\over\hbar}W(\varphi,\varphi^*)\}
\overleftarrow{\partial_a}=0.\eqno{(3.57)}$$

So, we have shown that the standard BV ansatz (3.56) follows from the
general functional representation (3.1), being the Darboux coordinates
and trivial measure density are chosen to work with.

On the other hand, the proposed functional integral (3.1) possesses, by
construction, a quite invariant and symmetric form.

\section{General Functional Integral of the Second Level}

In Section 3 we have formulated the unimodular involution relations
(3.5)-(3.7) for the hypergauge functions $G_a$. It is a remarkable
circumstance that these relations can be generated by means of a
unique supermechanism that synthesizes in itself the characteristic
features of the Hamiltonian and Lagrangian gauge--algebra--generating
equations.

In its own turn, the above--mentioned supermechanism will be shown
below to generate an effective action of the general functional
integral of the second level. Thus we shall make actually the first
step in hierarchical proliferation process that converts successively
the hypergauge Lagrangian multipliers into anticanonical pairs by
assigning a new antifield to each of the preceding--stage Lagrangian
multipliers.

To begin with, let us assign an antifield to each of the initial
Lagrangian multipliers:

$$\pi^a,\varepsilon(\pi^a)=\varepsilon_a,\quad\rightarrow\quad\pi^*_a,
\varepsilon(\pi^*_a)=\varepsilon_a+1. \eqno{(4.1)}$$

Let:

$$\Gamma^{\prime A^\prime}\equiv\left(\begin{array}{c}\Gamma^A\\\left(
\begin{array}{c}\pi^a\\\pi^*_a\end{array}\right)\end{array}\right),\quad
\varepsilon(\Gamma^{\prime A^\prime})\equiv\varepsilon^\prime_{A^\prime},
\quad A^\prime=1,\ldots,4N,\eqno{(4.2)}$$
be an extended set of the field--antifield variables.

Let us define the extended antisymplectic metric, antisymplectic
differential and antibrackets as follows:

$$E^{\prime A^\prime B^\prime}\equiv\left(\begin{array}{cc}E^{AB}&0\\
0&\left(\begin{array}{cc}0&\delta^{ab}\\-\delta^{ab}&0\end{array}\right)
\end{array}\right),\eqno{(4.3)}$$

$$\Delta^\prime\equiv{1\over2}(-1)^{{}^{\varepsilon^\prime
\hspace{-0.13cm}_{A^\prime}}}\hspace{-0.09cm}M^{-1}
\partial^\prime_{A^\prime}ME^{\prime A^\prime B^\prime}
\partial^\prime_{B^\prime},\eqno{(4.4)}$$

$$(A,B)^\prime\equiv A\overleftarrow{\partial^\prime_{C^\prime}}
E^{\prime C^\prime D^\prime}\overrightarrow{\partial^\prime_{D^\prime}}B.
\eqno{(4.5)}$$

The measure density $M(\Gamma)$ remains unchanged, so that the extended
measure is:

$$d\mu^\prime(\Gamma^\prime)\equiv d\pi d\pi^*d\mu(\Gamma).\eqno{(4.6)}$$

The extensions (4.3)-(4.6) obviously retain all the above--mentioned
formal properties of the antisymplectic differential and antibracket.

It is relevant at this stage to define the Planck parity
$\hbox{Pl}(A)$:

$$\hbox{Pl}(AB)=\hbox{Pl}(A)+\hbox{Pl}(B),\quad\hbox{Pl}(\hbar)\equiv1,
\eqno{(4.7)}$$

$$\hbox{Pl}(\Gamma^A)=0,\quad\hbox{Pl}(\pi^a)=-\hbox{Pl}(\pi^*_a)=1.
\eqno{(4.8)}$$

Next, let us consider the quantum master equation in its extended
version:

$$\Delta^\prime\exp\{{i\over\hbar}W^\prime(\Gamma^\prime)\}=0, \eqno{(4.9)}$$
under the extra conditions:

$$\hbox{Pl}\left(W^\prime(\Gamma^\prime)\right)=1,\quad
W^\prime(\Gamma^\prime)\bigl|_{\pi^*=0}=G_a(\Gamma)\pi^a,
\eqno{(4.10)}$$
where $G_a(\Gamma)$ are the first--level hypergauge functions considered above.

Let us seek for a solution to the problem (4.9), (4.10) in the form
of $\hbar$--power series expansion:

$$W^\prime(\Gamma^\prime)=\Omega+i\hbar\Xi+(i\hbar)^2\tilde{\Omega}+
\ldots\,\,. \eqno{(4.11)}$$
Then we find the following equations for the functions $\Omega$, $\Xi$,
$\tilde{\Omega}$:

$$(\Omega,\Omega)^\prime=0,\quad\hbox{Pl}(\Omega)=1,\quad
\Omega\bigl|_{\pi^*=0}=G_a\pi^a, \eqno{(4.13)}$$

$$(\Omega,\Xi)^\prime=\Delta\hspace{-0,05cm}^\prime\hspace{0.05cm}
\Xi,\quad\hbox{Pl}(\Xi)=0,\quad
\Xi\bigl|_{\pi^*=0}=0, \eqno{(4.14)}$$

$$(\Omega,\tilde{\Omega})^\prime=\Delta\hspace{-0,05cm}^\prime\hspace{0.05cm}
\Xi-{1\over2}(\Xi,\Xi)^\prime,
\quad\hbox{Pl}(\tilde{\Omega})=-1. \eqno{(4.15)}$$

In their own turn, these equations can be solved in the form of
$\pi,\pi^*$--power series expansions:

$$\Omega=G_a\pi^a-{1\over2}\pi^*_cU^c_{ab}\pi^b\pi^a(-1)^{\varepsilon_a}+
\ldots, \eqno{(4.16)}$$

$$\Xi=\pi^*_aV^a_b\pi^b+{1\over4}\pi^*_b\pi^*_aV^{ab}_{cd}\pi^d\pi^c
(-1)^{(\varepsilon_b+\varepsilon_c)}+\ldots, \eqno{(4.17)}$$

$$\tilde{\Omega}=\pi^*_a\tilde{G}{}^a+{1\over2}\pi^*_b\pi^*_a
\tilde{U}{}^{ab}_c\pi^c(-1)^{\varepsilon_b}+\ldots\,\,. \eqno{(4.18)}$$

We state that to the lowest order in $\pi,\pi^*$ the equations (4.13)-(4.15)
give exactly the unimodular involution relations (3.5)-(3.7), whereas
to higher $\pi,\pi^*$--orders one obtains all the compatibility conditions
for these relations. Thus the equations (4.13)--(4.15) appear to be
generating ones for the unimodular involution relations.

Now, let us consider the proposed general form of the Lagrangian
functional integral of the second level:

$$Z^\prime=\int\!\!\exp\{{i\over\hbar}[W(\Gamma)+W^\prime(\Gamma^\prime)+
G^\prime_a(\Gamma^\prime)\pi^{\prime a}]\}d\pi^\prime
d\mu^\prime(\Gamma^\prime), \eqno{(4.19)}$$
where: the first-level quantum action $W(\Gamma)$ satisfies the quantum
master equation (3.2); the second--level quantum action
$W^\prime(\Gamma^\prime)$ is defined above to be a solution of the problem
(4.9), (4.10);

$$G^\prime_a(\Gamma^\prime),\quad\varepsilon(G^\prime_a)\equiv
\varepsilon^\prime_a,
\quad a=1,\ldots,N, \eqno{(4.20)}$$
are new, second--level, hypergauge functions that satisfy the following
unimodular involution relations of the second level:

$$(G^\prime_a,G^\prime_b)=G^\prime_cU^{\prime c}_{ab}, \eqno{(4.21)}$$

$$\Delta^\prime G^\prime_a-U^{\prime b}_{ba}(-1)^{\varepsilon^\prime_b}=
G^\prime_bV^{\prime b}_a,\eqno{(4.22)}$$

$$V^{\prime a}_a=G^\prime_a\tilde{G}{}^{\prime a}, \eqno{(4.23)}$$

$$(W,G^\prime_a)^\prime=G^\prime_bX_a^{\prime b}, \eqno{(4.24)}$$

$$X^{\prime a}_a=G^\prime_a\bar{G}{}^{\prime a}. \eqno{(4.25)}$$
Besides, the matrix:

$$||{\partial
G^\prime_a\over\partial\pi^*_b}||\bigl|_{G^\prime=0}
\eqno{(4.26)}$$ is supposed to be even and nondegenerate, so that the gauge
equations $G^\prime_a=0$ are solvable with respect to the antifield variables
$\pi^*_a$.

The functional integral (4.19) is invariant under the following
BRST--type transformations:

$$\delta\Gamma^{\prime A^\prime}=(\Gamma^{\prime A^\prime},W-W^\prime+
G^\prime_a\pi^{\prime a})^\prime\mu^\prime, \eqno{(4.27)}$$

$$
\begin{array}{c}\delta\pi^{\prime a}=[-U^{\prime a}_{bc}\pi^{\prime c}
\pi^{\prime b}(-1)^{\varepsilon^\prime_b}-2X^{\prime a}_b\pi^{\prime b}+\\[9pt]
+2i\hbar(V^{\prime a}_b\pi^{\prime b}-\bar{G}{}^{\prime a})+2(i\hbar)^2
\tilde{G}{}^{\prime a}]\mu^\prime.\end{array}  \eqno{(4.28)}$$

Choosing the fermionic parameter $\mu^\prime$ to be the function:

$$\mu^\prime={i\over\hbar}\delta\Psi^\prime(\Gamma^\prime) \eqno{(4.29)}$$
that satisfies the equations:

$$i\hbar\Delta^\prime\delta\Psi^\prime-(W,\delta\Psi^\prime)^\prime=
G^\prime_a\delta K^{\prime a}, \eqno{(4.30)}$$

$$\Delta^\prime[G^\prime_a\delta K^{\prime a}+(W,\delta\Psi^\prime)^\prime]=0,
\eqno{(4.31)}$$
and making the additional variations:

$$\delta\Gamma^{\prime A^\prime}=(\Gamma^{\prime A^\prime},
\delta\Psi^\prime)^\prime,\quad\delta\pi^{\prime a}=\delta K^{\prime a},
\eqno{(4.32)}$$
one generates in the functional integral (4.19) canonical change
of the hypergauge functions $G^\prime_a$ alone:

$$G^\prime_a(\Gamma^\prime)\quad\rightarrow\quad
G^\prime_a\left(\Gamma^\prime+(\Gamma^\prime,\delta\Psi^\prime)^\prime\right).
\eqno{(4.33)}$$

Thus it is proven formally that the functional integral (4.19) does
not depend on the hypergauge variations (4.33). The same as in the
first--level case (3.15), the hypergauge variations (4.33) induce
the most general actual changes admitted by the relations (4.21)-(4.25)
for the second--level hypergauge surface $G^\prime_a=0$.

Choosing the trivial gauge:

$$G^\prime_a=\pi^*_a, \eqno{(4.34)}$$
and using the second condition (4.10),one returns to the first--level
functional
integral (3.1):

$$Z^\prime=Z \eqno{(4.35)}$$

Let us compare the general structure of the functional representations
(3.1) and (4.19). While in the expression (3.1) the variables $\pi^a$ are
nothing other but usual Lagrangian multipliers to the first--level
hypergauge functions $G_a$, in the expression (4.19) these field variables
acquire the corresponding antifields $\pi^*_a$, and these anticanonical pairs
appear to be working as a set of ``ghost'' variables with respect to the
second--level hypergauge $G^\prime_a$. It is a remarkable fact that the
classical ``ghost'' action $\Omega$, defined by the equations (4.13), (4.16),
possesses exactly the structure of the Hamiltonian BFV-generator, whereas the
first--level hypergauge functions $G_a$ play the role of the initially--given
first--class constraints. In their own turn, the new variables $\pi^\prime_a$
in
(4.19) are usual Lagrangian multipliers to the new, second--level, hypergauge
functions $G^\prime_a$.

The above--considered proliferation process can be continued by induction
for an arbitrary number of steps. At each step the former Lagrangian
multipliers acquire their antifields and become ghost variables with
respect to the hypergauges of the present stage. At each stage there is
no dependence of the functional integral on the admissible variations
of each hypergauge function entered.

\section{The Case of Field-Antifield Phase Space Reduced
by Second-Class Constraints}

While in the preceding Sections 3 and 4 we restricted ourselves by the
case of nondegenerate antisymplectic metric, now we have to comment in
brief the modifications needed in the degenerate metric case that corresponds
to the Dirac's counterpart of the above--considered formalism.

In more details, let the basic field--antifield variables (2.1) be
reduced originally by the second--class constraints (2.23), so that the
number of the required first--level hypergauge functions $G_a$ decreases
to become $N-L$. In that case all the above--given formulae retain their
true if one makes the following formal substitutions:

$$(A,B)\,\rightarrow\,(A,B)_{(D)},\quad\Delta\,\rightarrow\,\Delta_{(D)},\quad
d\mu(\Gamma)\,\rightarrow\,d\mu_{(D)}(\Gamma), \eqno{(5.1)}$$
where the definitions (2.27)-(2.31) are to be applied.

as an explicit example of the above--mentioned modifications, let us
give the Dirac's version of the first--level functional integral (3.1):

$$Z_{(D)}=\int\!\!\exp\{{i\over\hbar}[W(\Gamma)+G_a(\Gamma)\pi^a]\}d\pi
d\mu_{(D)}(\Gamma), \eqno{(5.2)}$$
where

$$\Delta_{(D)}\exp\{{i\over\hbar}W\}=0 \eqno{(5.3)}$$

$$(G_a,G_b)_{(D)}=G_cU^c_{ab}, \eqno{(5.4)}$$

$$\Delta_{(D)}G_a-U^b_{ba}(-1)^{\varepsilon_b}=G_bV^b_a, \eqno{(5.5)}$$

$$V^a_a=G_a\tilde{G}{}^a. \eqno{(5.6)}$$

The rest of the formulae should be modified in the same way.

Due to the presence of the constraint $\delta$--function $\delta(\Theta)$ in
the
integrand of (5.2), all the equalities (2.29), (2.30), (5.3)--(5.6) can be
weakened modulo a linear combination of second--class constraints.

\end{document}